\documentclass{revtex4}
\usepackage{graphicx}
\usepackage{fancyhdr}
\usepackage{amsmath,amssymb}
\usepackage{slashed}
\usepackage{epsfig,amsfonts,amsthm,subfigure,epstopdf}
\newcommand{\be}{\begin{equation}}
\newcommand{\ee}{\end{equation}}
\newcommand{\ba}{\begin{align}}
\newcommand{\ea}{\end{align}}
\newcommand{\bea}{\begin{eqnarray}}
\newcommand{\eea}{\end{eqnarray}}
\newcommand{\no}{\nonumber\\}

\usepackage{color}

\begin{document}

\title{Tree-level metastability bounds in two-Higgs doublet models}


\author{A. Barroso}
\email[E-mail: ]{barroso@cii.fc.ul.pt}
\affiliation{Centro de F\'{\i}sica Te\'{o}rica e Computacional,
    Faculdade de Ci\^{e}ncias,
    Universidade de Lisboa,
    Av.\ Prof.\ Gama Pinto 2,
    1649-003 Lisboa, Portugal}
\author{P.M.~Ferreira}
    \email[E-mail: ]{ferreira@cii.fc.ul.pt}
\affiliation{Instituto Superior de Engenharia de Lisboa - ISEL,
	1959-007 Lisboa, Portugal}
\affiliation{Centro de F\'{\i}sica Te\'{o}rica e Computacional,
    Faculdade de Ci\^{e}ncias,
    Universidade de Lisboa,
    Av.\ Prof.\ Gama Pinto 2,
    1649-003 Lisboa, Portugal}
\author{I. Ivanov}
\email[E-mail: ]{igor.ivanov@ulg.ac.be}
\affiliation{IFPA, Universit\'{e} de Li\`{e}ge, All\'{e}e du 6 Ao\^{u}t 17, b\^{a}timent B5a, 4000 Li\`{e}ge, Belgium}
\affiliation{Sobolev Institute of Mathematics, Koptyug avenue 4, 630090, Novosibirsk, Russia}
\author{Rui Santos}
    \email[E-mail: ]{rsantos@cii.fc.ul.pt}
\affiliation{Instituto Superior de Engenharia de Lisboa - ISEL,
	1959-007 Lisboa, Portugal}
\affiliation{Centro de F\'{\i}sica Te\'{o}rica e Computacional,
    Faculdade de Ci\^{e}ncias,
    Universidade de Lisboa,
    Av.\ Prof.\ Gama Pinto 2,
    1649-003 Lisboa, Portugal}

\begin{abstract}
The two Higgs doublet model has a rich vacuum structure, including the possibility
of existence of two Standard Model-like minima at tree-level. It is therefore possible
that the universe's vacuum is metastable, and a deeper minimum exists. We present the
analytical conditions one must demand of the potential's parameters to prevent that
possibility, and analyse what the current LHC data tells us about the eventual existence of
that second minimum.
\end{abstract}

\maketitle

\thispagestyle{fancy}

%%%%%%%%%%%%%%%%%%%%%%%%%%%%%%%%%%
\section{Introduction}

The two Higgs doublet model (2HDM)~\cite{Lee:1973iz} is one of the simplest extensions of the Standard
Model (SM) of particle physics, in which the number of scalar doublets is doubled compared to the
SM. This simple addition makes for a richer scalar spectrum, which includes two CP-even scalars, the lightest $h$ and the heaviest $H$, a pseudoscalar, $A$, and a charged scalar, $H^\pm$.
The 2HDM boasts very interesting phenomenology, including possible spontaneous CP violation,
tree-level flavour changing neutral currents mediated by scalars and dark matter candidates. For
a recent review, see~\cite{Branco:2011iw}. The recent discovery of the Higgs boson~\cite{:2012gk,:2012gu}
enables us to further constrain the 2HDM parameter space, and ascertain whether the model
survives comparison with data. In fact, the 2HDM does a very good job describing the LHC
results~\cite{Chen:2013kt,Belanger:2012gc,Chang:2012ve,Ferreira:2011aa}.

In the SM there is only one possible type of vacuum. In the 2HDM, on the other hand, charge
breaking vacua may occur, and in those situations the photon would acquire a mass. Also, as
was already mentioned, we also have possible minima in which the CP symmetry is spontaneously
broken, alongside with the electroweak gauge symmetry. But a remarkable property of the
2HDM potential is that these different types of minima cannot
coexist~\cite{Ferreira:2004yd,Barroso:2005sm}. We call a vacuum which
breaks electroweak symmetry but preserves the electromagnetic and CP symmetries a ``normal" minimum.
And whenever a normal minimum exists, any possible charge or CP breaking stationary points are
{\em necessarily} saddle points, and lie above the normal minimum. The stability of the normal
minimum against charge or CP breaking is thus guaranteed. But there is still another difference
in the 2HDM vacuum structure regarding the SM: for certain choices of parameters, the 2HDM potential
can have {\em two} normal minima simultaneously~\cite{Barroso:2007rr,Ivanov:2006yq,Ivanov:2007de}.
The two Higgs doublets, $\Phi_1$ and $\Phi_2$, would have vacuum expectation values (vevs)
$v_1$ and $v_2$ such that $v_1^2 + v_2^2\, =\, 246$ GeV$^2$ in one of the minima, and all elementary
particles would have the known masses - this would be the minimum where the universe is currently at.
In the second minimum, however - and this minimum could be deeper or higher than ours - the fields
would have vevs  $\{v^\prime_1\,,\,v^\prime_2\}$, with ${v^\prime}^2_1\,+\,{v^\prime}^2_2\,\neq
246$ GeV$^2$ - and the elementary particles might have masses much smaller, or larger, than
what is observed.

It is therefore possible that ``our" vacuum is not the state of lowest energy of the theory. Thus
the universe would be in a metastable state, with the possibility of tunneling to the
deeper vacuum. We call this situation the ``panic vacuum". In~\cite{Barroso:2012mj} we presented the
conditions that the parameters of the potential need to obey so one can avoid the presence of a panic vacuum in the softly broken Peccei-Quinn~\cite{Peccei:1977hh} version of the 2HDM. We also showed that the LHC data
already excludes most of the parameter space where panic vacua might occur in this model. In this talk
I will discuss the existence of double neutral minima in the 2HDM potential with a $Z_2$ discrete
symmetry. That is the most used version of the 2HDM, and panic vacua do occur in it, for plenty of
the model's parameter space. The bounds one has to impose on the potential to avoid such minima
were deduced in~\cite{disc} and will be reviewed here, as well as what the current LHC
data tells us about the existence of the second minimum.

\section{The vacuum structure of the 2HDM}

The most general 2HDM potential has 14 real parameters, which can be reduced to 11 using
the reparametrization invariances of the model. That model, however, leas to tree-level
flavour changing neutral currents mediated by scalars. In order to avoid them - since they
are incredibly constrained by experimental measurements - one usually imposes a discrete
$Z_2$ symmetry on the lagrangian, such that $\Phi_1\rightarrow \Phi_1$ and
$\Phi_2\rightarrow -\Phi_2$. This was first proposed by Glashow, Weinberg and
Paschos~\cite{Glashow:1976nt,Paschos:1976ay}, and the resulting scalar potential has only
seven independent real parameters. Its parameter space is however extremely constrained, which
is why one many times adds a softly breaking term $m_{12}$. The softly broken potential is
therefore written as
\bea
V &=&
m_{11}^2 |\Phi_1|^2
+ m_{22}^2 |\Phi_2|^2
- m_{12}^2 \left(  \Phi_1^\dagger \Phi_2 + h.c. \right)
\no & &
+ \frac{1}{2} \lambda_1 |\Phi_1|^4
+ \frac{1}{2} \lambda_2 |\Phi_2|^4
+ \lambda_3 |\Phi_1|^2 |\Phi_2|^2
+ \lambda_4 |\Phi_1^\dagger\Phi_2|^2
+
\frac{1}{2} \lambda_5 \left[\left( \Phi_1^\dagger\Phi_2 \right)^2
+ h.c. \right],
\label{eq:pot}
\eea
where we have also imposed CP symmetry on the scalar potential and all the parameters shown
are real. As we have already mentioned, the 2HDM can have charge breaking vacua, wherein
the doublets acquire vevs $\langle\Phi_1 \rangle_{CB} = ( 0 \, , \, c_1)^T/\sqrt{2}$ and
$\langle\Phi_2 \rangle_{CB} = ( c_2 \, , \, c_3)^T/\sqrt{2}$. If, simultaneously, there is
a ``normal" solution $(v_1\,,\,v_2)/\sqrt{2}$ of the minimization equations, the depth of
the potential at the charge breaking stationary point, $V_{CB}$, is related to the
depth of the potential at the normal stationary point, $V_N$, by~\cite{Ferreira:2004yd,Barroso:2005sm}
\be
V_{CB} - V_N\,=\,\left(\frac{m^2_{H^\pm}}{4 v^2}\right)_N\,
\left[(v_1 c_3 - v_2 c_1)^2 + v_1^2 c_2^2\right]\, ,
\label{eq:diffcb}
\ee
where $m^2_{H^\pm}$ is the square of the charged scalar mass at the normal stationary point -
meaning, if that is a minimum we will have $V_{CB} - V_N > 0$; further, it was also shown that
the charge breaking stationary point is, in this case, necessarily a saddle point. {\em Ergo},
the normal minimum is the global one, and no tunneling to a deeper minimum can occur. Likewise,
if there is a CP breaking stationary point, with vevs such as $\langle\Phi_1 \rangle_{CP} =
( 0 \, , \, \bar{v}_1)^T/\sqrt{2}$ and $\langle\Phi_2 \rangle_{CP} = ( 0 \, , \, \bar{v}_2 e^{i\theta})^T/\sqrt{2}$, the difference of depths of the potential at this stationary point
and a normal one is given by
\be
V_{CP} - V_N\,=\,\left(\frac{m^2_A}{4 v^2}\right)_N\,
\left[(\bar{v}_2 v_1 \cos\theta - \bar{v}_1 v_2)^2 + \bar{v}_2^2 v_1^2 \sin^2\theta\right]\, ,
\label{eq:diffcp}
\ee
where $m^2_A$ is the pseudoscalar mass at the normal stationary point. Again, if the normal
stationary point is a minimum, it will be the deepest and thus stable against tunneling.
In short, the existence of a normal minimum guarantees that the global minimum of the potential
is also normal. However, the 2HDM, due to the soft breaking term, can have two normal minima~\cite{Barroso:2007rr,Ivanov:2006yq,Ivanov:2007de}. Further, it may be shown that if the
potential has a depth equal to $V_N$ for the minimum with vevs $\{v_1\,,\,v_2\}$, and a depth
$V_{N^\prime}$ for the minimum with vevs $\{v^\prime_1\,,\,v^\prime_2\}$, we have
\begin{eqnarray}
V_{N^\prime} - V_{N} &=& \frac{1}{4}\,\left[\left(\frac{m^2_{H^\pm}}{v^2}\right)_{N}
- \left(\frac{m^2_{H^\pm}}{v^2}\right)_{N^\prime}\right]\,
(v_1 v^\prime_2 - v_2 v^\prime_1)^2 \nonumber \\
 &=& \frac{m^2_{12}}{4 v_1 v_2}\,
\left(1 - \frac{v_1 v_2 }{v^\prime_1 v^\prime_2}\right)\,
(v_1 v^\prime_2 - v_2 v^\prime_1)^2\, .
\end{eqnarray}
In this expression, both the charged Higgs mass $m^2_{H^\pm}$ and the sum of the squared vevs $v^2$ is
evaluated at each minima. It is not therefore clear which is the deepest minimum. But in~\cite{Ivanov:2006yq,Ivanov:2007de,ivanovPRE} the conditions for the existence
of two neutral minima in the 2HDM were established. They were cast into a simpler form
in~\cite{Barroso:2012mj} for the Peccei-Quinn model, and in~\cite{disc} for the $Z_2$ one.
In~\cite{disc} we also presented a thorough deduction of the bounds to avoid panic vacua, and
their generalisation for all 2HDM CP-conserving potentials. Therein we were able to deduce
simple necessary conditions for the existence of two normal minima for any such potentials,
which we will not reproduce here for briefness.
Fortunately, the study of panic vacua does not require that one analyses whether or not
 the potential satisfies those conditions. In fact, all we need do is compute the
following quantity $D$, which we call a discriminant:
\be
D \,=\, m^2_{12} (m^2_{11} - k^2 m^2_{22}) (\tan\beta - k)\hspace{5mm}
\textrm{with}
\hspace{5mm}
k = \sqrt[4]{\frac{\lambda_1}{\lambda_2}}.
\label{eq:disc}
\ee
As usual, $\tan\beta = v_2/v_1$, written with the
vevs of ``our" vacuum. The existence of a panic vacuum is thus summarised in the
following theorem:
\be
\mbox{
{\em Our vacuum is the global minimum of the potential if and only if $D > 0$.}
}
\label{eq:cond}
\ee

It is therefore extremely simple to ascertain whether the 2HDM minimum is, or is not,
the global minimum of the model: all one has to do is to compute the value of $D$ above,
having reconstructed the parameters of the model from experiments (this may prove
to be difficult, but it is, in principle, quite achievable). Up to date, there is no
evidence of the existence of any scalars beyond that which is predicted by the SM.
Nonetheless, as we will now see, the current LHC data already allow us to exclude most
of the parameter space where panic vacua might occur.

\section{Panic vacua exclusion using LHC data}

The 2HDM scalar potential we are considering has 8 independent parameters. That
is a vast parameter space, but it can be constrained in many different ways. In the first
place it needs to have one minimum with vevs which originate the correct masses for the $W$
and $Z$ bosons, {\em i.e.} the doublets must need have vevs $v_1$ and $v_2$ such that
$v_1^2 + v_2^2\, =\, 246$ GeV$^2$. Further, since the discovery of the Higgs boson, we
must ensure that the lightest CP-even scalar has a mass of 125 GeV. On the other hand,
the potential has to be bounded from below, which imposes the following constraints
on its quartic couplings:
\begin{eqnarray}
\lambda_1 > 0 & , &  \lambda_2 > 0 \; ,\nonumber \\
\lambda_3 > -\sqrt{\lambda_1 \lambda_2} & , &
\lambda_3 + \lambda_4 - |\lambda_5| > -\sqrt{\lambda_1 \lambda_2} \;.
\label{eq:bfb}
\end{eqnarray}
Then, just as in the SM, we need to ensure that the model obeys perturbative
unitarity, which again constrains the potential's quartic
couplings~\cite{Kanemura:1993hm,Akeroyd:2000wc}. And of course, the model also
must comply with the precision electroweak constraints, via the bounds on the
S, T and U parameters~\cite{Peskin:1991sw,lepewwg,gfitter1,gfitter2}. However,
all of these bounds apply exclusively to the scalar sector of the theory, but one
must be aware that there are plenty of restrictions on extended scalar sectors
from their interactions with fermions, namely from B-physics experiments.  The $Z_2$
symmetry imposed on the scalar potential is extended to the fermion sector in
a variety of ways, we will consider only the two most studied: in model Type I
the fermion fields transform under the global $Z_2$ in such a way that only $\Phi_2$ couples
to the fermions; and in model Type II $\Phi_2$ couples only to up-type quarks, and $\Phi_1$ to
down-type quarks and charged leptons. Both of these models possess very different phenomenologies,
and B-physics bounds have been taken into account in our simulations~\cite{bphys1,bphys2}.
With all of these constraints in place, we performed a vast scan of the 2HDM parameter
space, for both models considered, taking $m_h = 125$ GeV, $125 < m_H < 900$ GeV,
$90 < m_A, m_{H^\pm} < 900$ GeV, $-\pi/2 < \alpha < \pi/2$, $1 < \tan\beta < 40$ and
$|m^2_{12}| < 900$ GeV$^2$. At this point we are ready to compute the observables which
will be compared to the LHC data.

In order to reduce the large uncertainties present in the calculation of hadronic
cross sections at the LHC, we consider the ratios between the
observed rates of the Higgs boson decaying into certain particles and their expected
SM values. If we assume that what is being observed is described by the 2HDM, that ratio is
defined, for a given final decay state $f$ of the lightest CP-even
Higgs boson $h$, as
\be
R_f \,=\, \frac{\sigma^{2HDM} (pp\rightarrow h) \,BR^{2HDM}(h \rightarrow f)}
{\sigma^{SM} (pp\rightarrow h) \,BR^{SM}(h \rightarrow f)}.
\ee
Thus both production cross sections $\sigma$ and branching ratios (BR) of the Higgs boson
are included, and we have considered all possible Higgs production mechanisms at the
LHC.

\begin{figure}[ht]
\centering
\includegraphics[height=6cm,angle=0]{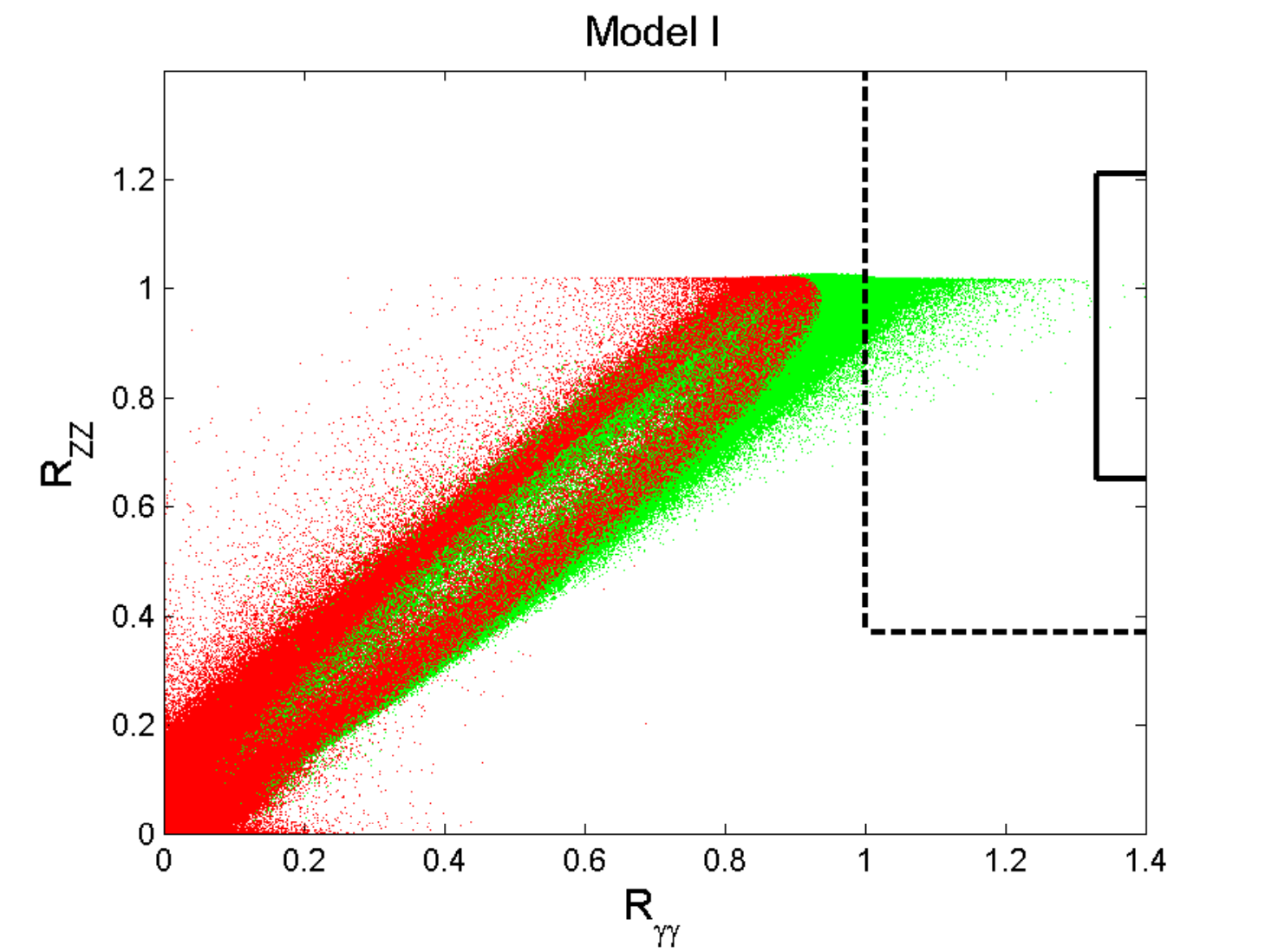}
\includegraphics[height=6cm,angle=0]{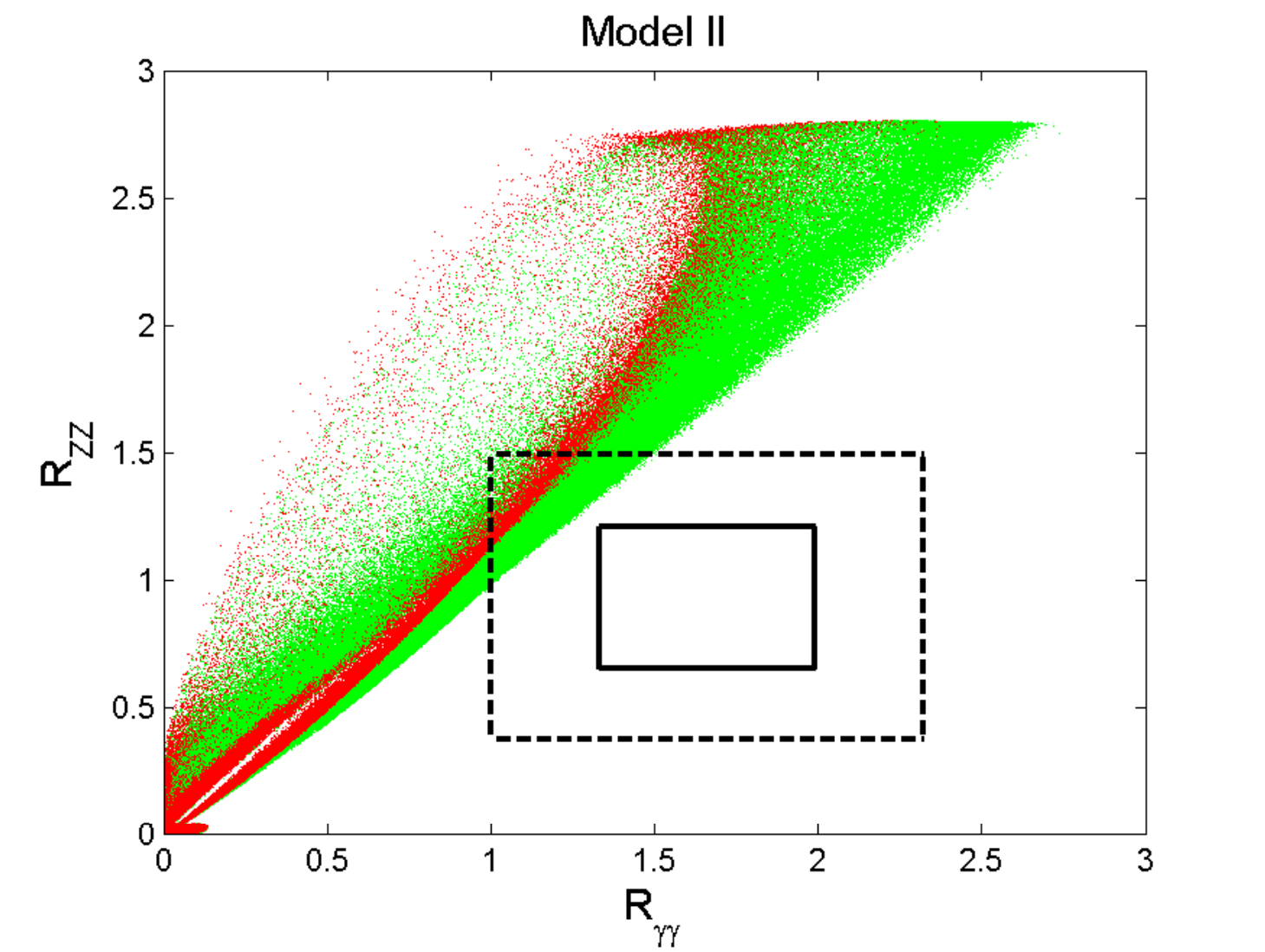}
\caption{$R_{ZZ}$ as a function of $R_{\gamma\gamma}$ for models I and II. Green
(light) points represent all points generated, red (dark) ones correspond to
the existence of a panic vacuum ($D < 0$). The solid (dashed) lines correspond to
$1\sigma$ ($2\sigma$) experimental bounds on the rates.}
\label{fig:zph}
\end{figure}

In Fig.~\ref{fig:zph} we present, for both models under consideration, the
rates of the light Higgs $h$ into two $Z$ bosons {\em versus} the rate
of $h$ into two photons. In green (light grey) are all the points obtained in
our scan, with the above constraints, both theoretical and experimental. In red (black)
we show the points for which a panic vacuum occurs - meaning, points for which the value of $D$,
calculated from eq.~\eqref{eq:disc}, is negative.
It is important to consider that the density of points generated is so large that there are
many green points scattered in the middle of the red ones - meaning, the areas marked red are not necessarily excluded (a consequence, of course, of the fact that we are dealing with an 8-dimensional
parameter space, and these figures are only 2-dimensional). Fig.~\ref{fig:zph}
clearly shows that there are regions, in the plane
 $R_{\gamma\gamma}$-$R_{ZZ}$, which are completely free from panic vacua.
The solid and dashed lines shown in the plots correspond to conservative $1\sigma$
and $2\sigma$ intervals on the combined values for $R_{ZZ}$ and $R_{\gamma\gamma}$,
$R_{ZZ} = 0.93 \pm 0.28$, $R_{\gamma\gamma} = 1.66 \pm 0.33$, which we took from
ref.~\cite{average}, based on the LHC data before the Moriond conference~\cite{cms_atlas}.
Notice that after the recent Moriond update on the LHC
results~\cite{Moriond} these numbers may have changed substantially, but there isn't yet
an official combination of the ATLAS and CMS results. However, the plots we show
in this communication have the advantage of being easily adapted for changing LHC
experimental bounds, by simply drawing over them different black lines.

The remarkable thing is how much the current LHC data already can tell us about the
nature of the 2HDM vacuum, even if no extra scalars have been found. In fact, as can be
seen from Fig.~\ref{fig:zph}, the panic points are distant from even
the $2\sigma$ bands, which include some non-panic region as well, for model Type I.
That does not occur in model Type II, in which
some of the panic region is inside the $2\sigma$ region. But notice that
there are many choices of parameter space values still allowed by the current
data for Model II which do not lead to panic vacua. Thus, at least in these two variables, both
types of models are capable of describing the current data. Nonetheless, that data does not exclude the possibility, in model type II, of our vacuum being metastable.

\begin{figure}[ht]
\centering
\includegraphics[height=6cm,angle=0]{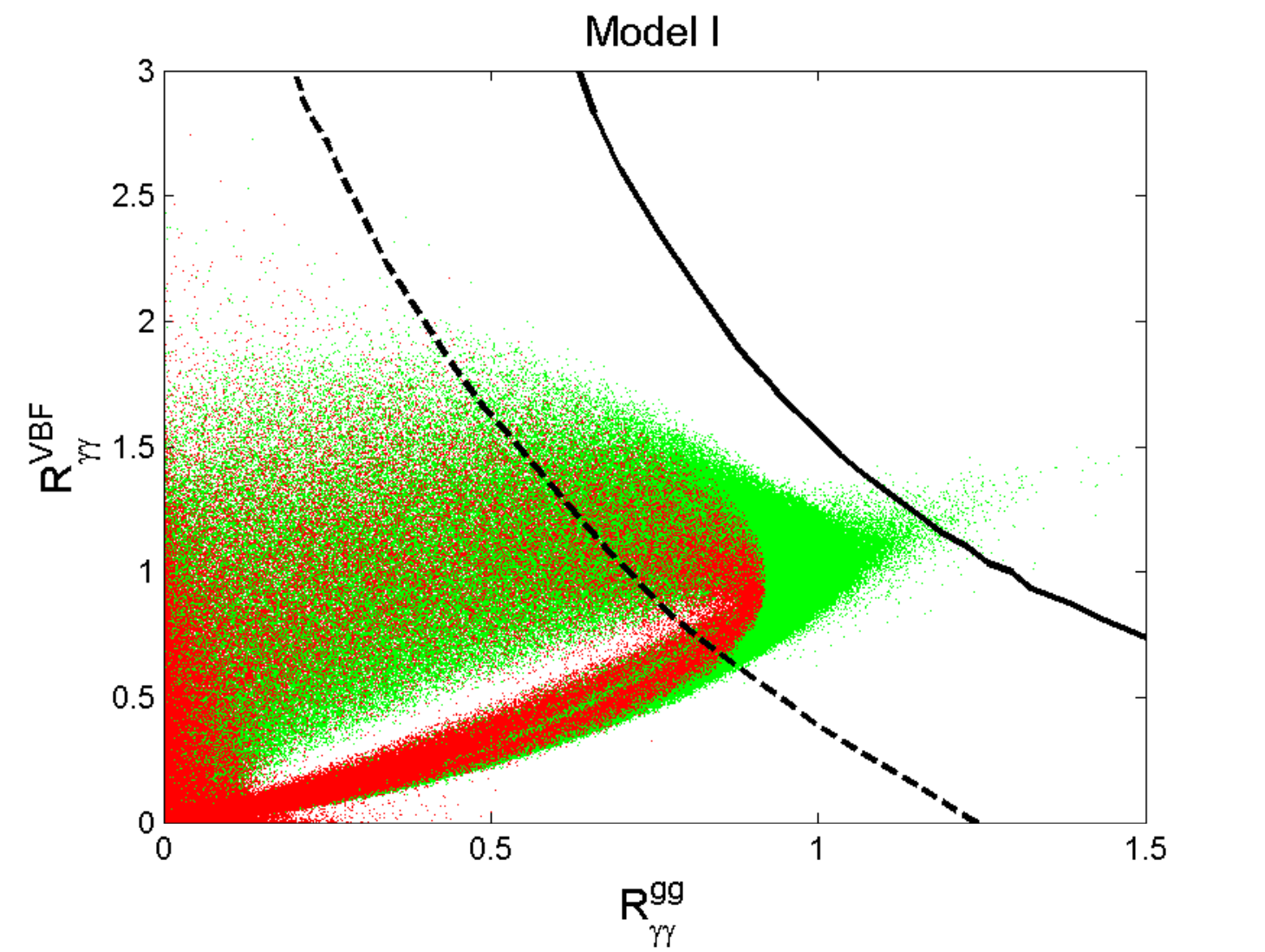}
\includegraphics[height=6cm,angle=0]{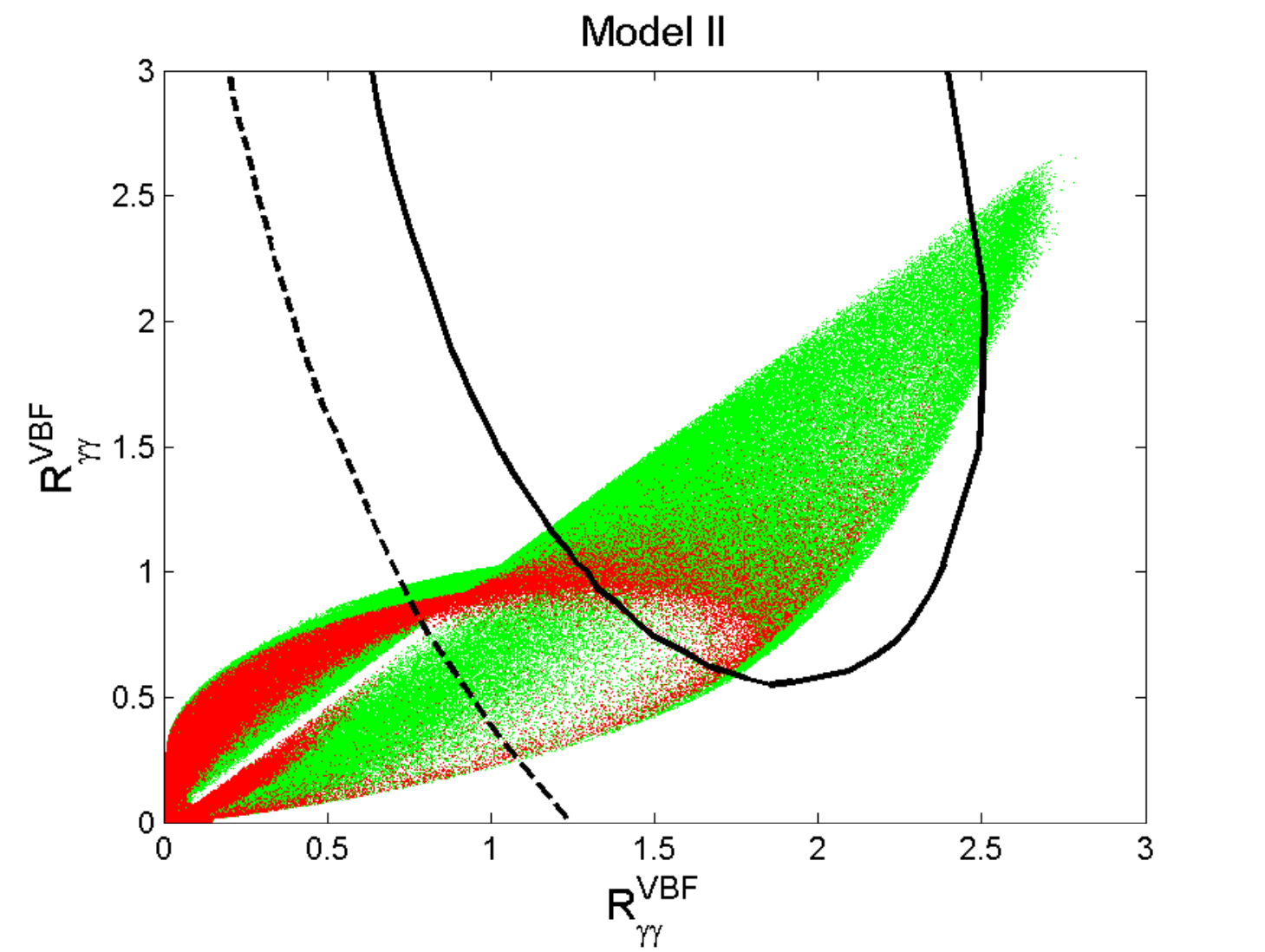}
\caption{$R_{\gamma\gamma}$, with Higgs production via gluon-gluon fusion {\em versus}
vector boson fusion production, for models I and II. Same colour codes as Fig.~\ref{fig:zph}.}
\label{fig:ggvbf}
\end{figure}

It has been possible to measure at the LHC - with considerable uncertainty - the production
of Higgs bosons via different processes, namely gluon-gluon fusion and vector boson fusion (VBF).
Analysing these processes separately gives us information about the coupling of the Higgs to both
fermion (the gluon-gluon process) and gauge bosons (the VBF one). We shall use the results of the ATLAS
experiment~\cite{cms_atlas}, which appear as $1\sigma$ and $2\sigma$ ellipses in the
$R^{gg}_{\gamma\gamma}$-$R^{VBF}_{\gamma\gamma}$ plane. Our results appear in
Fig.~\ref{fig:ggvbf}, for both Type I and Type II models. We observe that the experimental exclusion
of points with panic vacua is not as thorough as that which happened with the
previous observables. In model Type I it is not possible to exclude, at $2\sigma$,
the existence of panic vacua. However,
the panic vacua points which now seem allowed have been excluded in Fig.~\ref{fig:zph}.
For model Type II, even the $1\sigma$ bands include panic vacua solutions. We observe,
nonetheless, that the ellipses contain plenty of green/light grey points as well - which means
that there are many allowed choices of parameters for which panic vacua do not occur.
We even see that in these
variables the Type II model agrees with the data at the $1\sigma$ level, something which the Type I model
cannot achieve.

\begin{figure}[ht]
\centering
\includegraphics[height=6cm,angle=0]{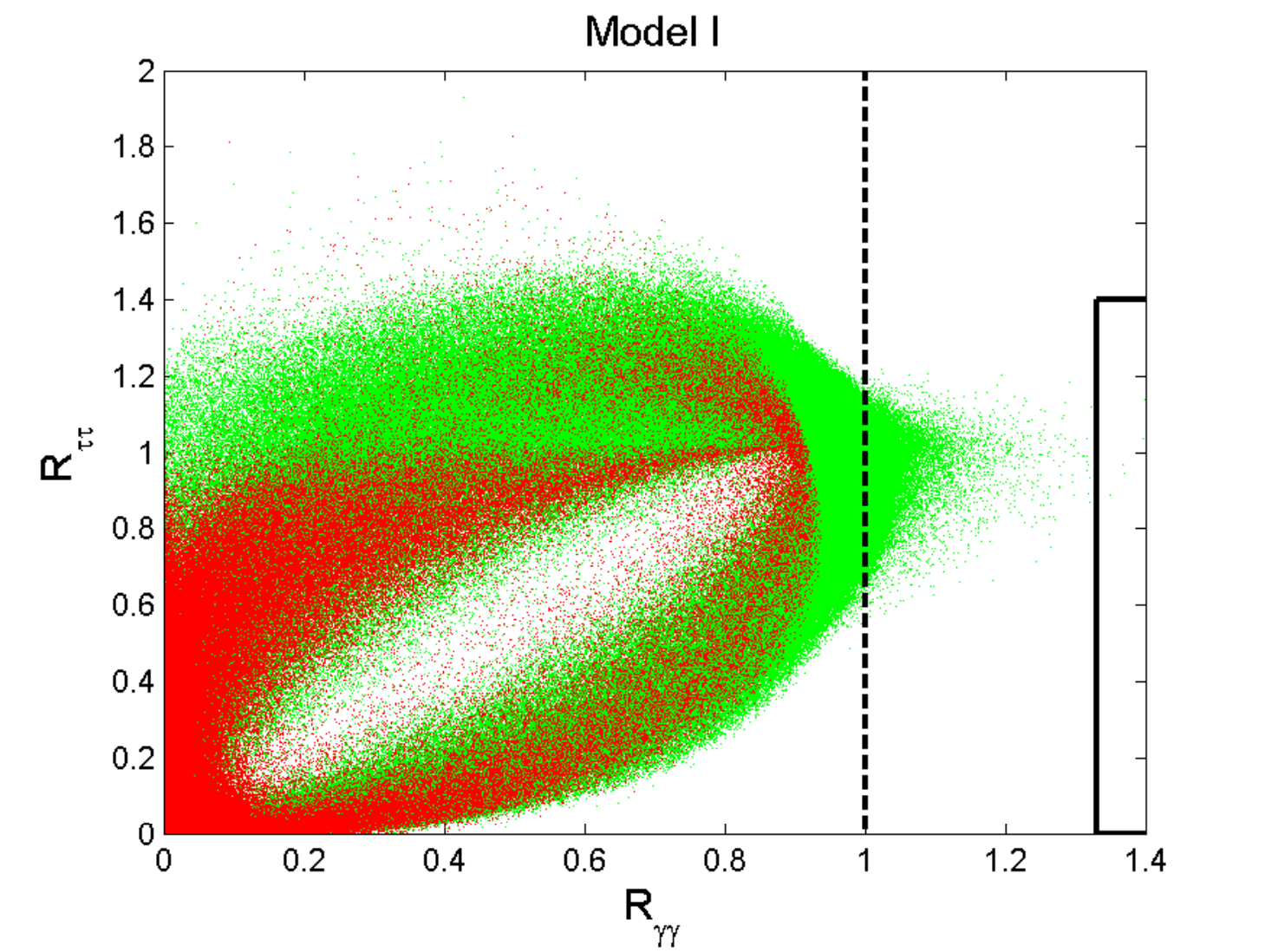}
\includegraphics[height=6cm,angle=0]{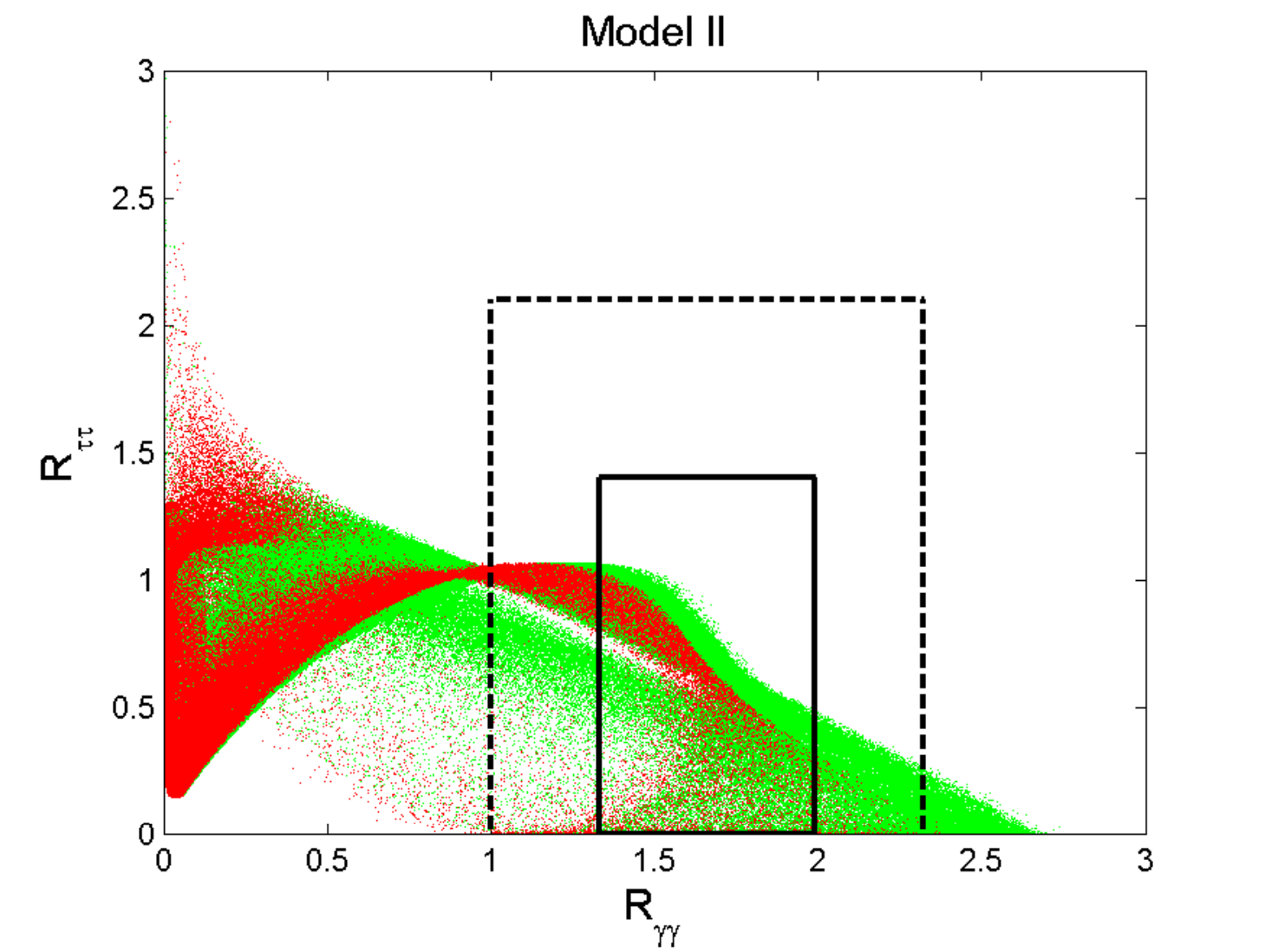}
\caption{$R_{\tau\tau}$ as a function of $R_{\gamma\gamma}$ for models Type I and Type II.
Same colour codes as Fig.~\ref{fig:zph}}
\label{fig:tauph}
\end{figure}

The current results for a Higgs decaying to $\tau\bar{\tau}$ are compatible with the expected
SM value. In fact, ATLAS measured $R_{\tau\tau} = 0.7 \pm 0.7$ and CMS
$R_{\tau\tau} = 0.72 \pm 0.52$. We can see how the panic vacua points
are distributed in the $\{R_{\tau\tau}\,,\,R_{\gamma\gamma}\}$ in Fig.~\ref{fig:tauph}.
The $\tau\tau$ data (we represent the less restrictive bounds, those of ATLAS),
tells us that panic vacua are
$2\sigma$ disfavored in model Type I, and that model barely agrees, at $2\sigma$, with the LHC results in $R_{\gamma\gamma}$, agreeing at $1\sigma$ in $R_{\tau\tau}$).
In model Type II, there are many panic vacua solutions
not excluded by the data at $1\sigma$; but for much of Model II's parameter space, we have
agreement with the experimental results at the $1\sigma$ level, with or without panic
vacua.

\section{Conclusions}

The rich vacuum structure of the 2HDM includes the possibility of two neutral minima
being able to coexist at different depths. Thus there is the possibility that
the vacuum we are currently living in is metastable, which we called the panic vacuum.
It is possible to find an
extremely simple criterium, the discriminant $D$ of eq.~\eqref{eq:disc} being positive,
which, if obeyed, guarantees that no metastability occurs. We emphasise that this situation
is quite different from the SM one: there a metastable (or even unstable) vacuum may develop
but only due to radiative corrections. In fact, the importance of radiative corrections to the
results found here cannot be overstated, and remains an open question. We also performed an estimate of the lifetime of the false vacuum in the panic situation, and verified that it is, for the vast majority of the 
points scanned, much inferior to the age of the universe - as such, these panic vacua are indeed to be
excluded, since ``our" vacuum would have decayed long ago. Nevertheless, we see
that the current LHC results already tell us a lot about
the stability of the vacuum in the 2HDM. For instance, a measurement of $R_{ZZ}$
and $R_{\gamma\gamma}$ very close to 1, with sufficient precision, would exclude the possibility
of panic vacua. We also saw that the values of the
parameters of the potential which produce panic vacua do not correspond to uninteresting
regions of the model - rather, they predict observables which are not absurd
and indeed may fall into the current experimental bounds. This, by itself, indicates
the need to take seriously this possibility of vacuum instability in the 2HDM.

\begin{acknowledgments}
The works of A.B., P.M.F. and R.S. are supported in part by the Portuguese
\textit{Funda\c{c}\~{a}o para a Ci\^{e}ncia e a Tecnologia} (FCT)
under contract PTDC/FIS/117951/2010, by FP7 Reintegration Grant, number PERG08-GA-2010-277025,
and by PEst-OE/FIS/UI0618/2011.
%The work of J.P.S. is funded by FCT through the projects
%CERN/FP/109305/2009 and  U777-Plurianual,
%and by the EU RTN project Marie Curie: PITN-GA-2009-237920.
I.P.I. is thankful to CFTC, University of Lisbon, for their hospitality.
His work is supported by grants RFBR 11-02-00242-a,
RF President grant for scientific schools NSc-3802.2012.2, and the
Program of Department of Physics SC RAS and SB RAS "Studies of Higgs boson and exotic particles at LHC".
\end{acknowledgments}

\bigskip % extra skip inserted

\begin{thebibliography}{99} % Use for 10-99 references
\bibitem{Lee:1973iz}
 T.~D.~Lee,
 %``A Theory of Spontaneous T Violation,''
 Phys.\ Rev.\  D {\bf 8} (1973) 1226.
 %%CITATION = PHRVA,D8,1226;%%

\bibitem{Branco:2011iw}
  G.~C.~Branco, P.~M.~Ferreira, L.~Lavoura, M.~N.~Rebelo, M.~Sher and J.~P.~Silva,
  %``Theory and phenomenology of two-Higgs-doublet models,''
  Phys.\ Rept.\  {\bf 516}, 1 (2012)
  [arXiv:1106.0034 [hep-ph]].

\bibitem{:2012gk}
G.~Aad {\it et al.}  [ATLAS Collaboration],
  %``Observation of a new particle in the search for the Standard Model Higgs
  %boson with the ATLAS detector at the LHC,''
Phys.\ Lett.\ B \textbf{716}, 1 (2012)
[arXiv:1207.7214 [hep-ex]].

\bibitem{:2012gu}
S.~Chatrchyan \textit{et al.}  [CMS Collaboration],
  %``Observation of a new boson at a mass of 125 GeV with the CMS experiment
  %at the LHC,''
Phys.\ Lett.\ B \textbf{716}, 30 (2012)
[arXiv:1207.7235 [hep-ex]].

%\cite{Chen:2013kt}
\bibitem{Chen:2013kt}
 C.~-Y.~Chen and S.~Dawson,
 %``Exploring Two Higgs Doublet Models Through Higgs Production,''
 arXiv:1301.0309 [hep-ph].
 %%CITATION = ARXIV:1301.0309;%%

%\cite{Belanger:2012gc}
\bibitem{Belanger:2012gc}
 G.~Belanger, B.~Dumont, U.~Ellwanger, J.~F.~Gunion and S.~Kraml,
 %``Higgs Couplings at the End of 2012,''
 arXiv:1212.5244 [hep-ph].
 %%CITATION = ARXIV:1212.5244;%%

%\cite{Chang:2012ve}
\bibitem{Chang:2012ve}
 S.~Chang, S.~K.~Kang, J.~-P.~Lee, K.~Y.~Lee, S.~C.~Park and J.~Song,
 %``Comprehensive study of two Higgs doublet model in light of the new boson
with mass around 125 GeV,''
 arXiv:1210.3439 [hep-ph].
 %%CITATION = ARXIV:1210.3439;%%

%\cite{Ferreira:2011aa}
\bibitem{Ferreira:2011aa}
 P.~M.~Ferreira, R.~Santos, M.~Sher and J.~P.~Silva,
 %``Implications of the LHC two-photon signal for two-Higgs-doublet models,''
 Phys.\ Rev.\ D {\bf 85}, 077703 (2012)
 [arXiv:1112.3277 [hep-ph]].
 %%CITATION = ARXIV:1112.3277;%%

\bibitem{disc}
A.~Barroso, P.~M.~Ferreira, I.~P.~Ivanov and R.~Santos,
  %``Metastability bounds on the two Higgs doublet model,''
  arXiv:1303.5098 [hep-ph].

\bibitem{Ferreira:2004yd}
 P.~M.~Ferreira, R.~Santos and A.~Barroso,
 %``Stability of the tree-level vacuum in two Higgs doublet models against
 %charge or CP spontaneous violation,''
 Phys.\ Lett.\  B {\bf 603} (2004) 219
 [Erratum-ibid.\  B {\bf 629} (2005) 114]
 [arXiv:hep-ph/0406231].
 %%CITATION = PHLTA,B603,219;%%

\bibitem{Barroso:2005sm}
 A.~Barroso, P.~M.~Ferreira and R.~Santos,
 %``Charge and CP symmetry breaking in two Higgs doublet models,''
 Phys.\ Lett.\  B {\bf 632} (2006) 684
 [arXiv:hep-ph/0507224].
 %%CITATION = PHLTA,B632,684;%%

\bibitem{Barroso:2007rr}
 A.~Barroso, P.~M.~Ferreira and R.~Santos,
 %``Neutral minima in two-Higgs doublet models,''
 Phys.\ Lett.\  B {\bf 652} (2007) 181
 [arXiv:hep-ph/0702098].
 %%CITATION = PHLTA,B652,181;%%

\bibitem{Ivanov:2006yq}
 I.~P.~Ivanov,
 %``Minkowski space structure of the Higgs potential in 2HDM,''
 Phys.\ Rev.\  D {\bf 75} (2007) 035001
 [Erratum-ibid.\  D {\bf 76} (2007) 039902]
 [arXiv:hep-ph/0609018].
 %%CITATION = PHRVA,D75,035001;%%

\bibitem{Ivanov:2007de}
 I.~P.~Ivanov,
 %``Minkowski space structure of the Higgs potential in 2HDM. II. Minima,
 %symmetries, and topology,''
 Phys.\ Rev.\  D {\bf 77} (2008) 015017
 [arXiv:0710.3490 [hep-ph]].
 %%CITATION = PHRVA,D77,015017;%%

\bibitem{Barroso:2012mj}
  A.~Barroso, P.~M.~Ferreira, I.~P.~Ivanov, R.~Santos and J.~P.~Silva,
  %``Evading death by vacuum,''
  arXiv:1211.6119 [hep-ph].
  %%CITATION = ARXIV:1211.6119;%%

\bibitem{Peccei:1977hh}
 R.~D.~Peccei and H.~R.~Quinn,
 %``CP Conservation in the Presence of Instantons,''
 Phys.\ Rev.\ Lett.\  {\bf 38} (1977) 1440.
 %%CITATION = PRLTA,38,1440;%%

\bibitem{Glashow:1976nt}
 S.~L.~Glashow and S.~Weinberg,
 %``Natural Conservation Laws for Neutral Currents,''
 Phys.\ Rev.\  D {\bf 15} (1977) 1958.
 %%CITATION = PHRVA,D15,1958;%%

\bibitem{Paschos:1976ay}
 E.~A.~Paschos,
 %``Diagonal Neutral Currents,''
 Phys.\ Rev.\  D {\bf 15} (1977) 1966.
 %%CITATION = PHRVA,D15,1966;%%

\bibitem{ivanovPRE}
  I.~P.~Ivanov,   Phys.\ Rev.\ E {\bf 79}, 021116 (2009).

\bibitem{Kanemura:1993hm}
 S.~Kanemura, T.~Kubota and E.~Takasugi,
 %``Lee-Quigg-Thacker bounds for Higgs boson masses in a two doublet model,''
 Phys.\ Lett.\  B {\bf 313} (1993) 155
 [arXiv:hep-ph/9303263].
 %%CITATION = PHLTA,B313,155;%%

\bibitem{Akeroyd:2000wc}
  A.~G.~Akeroyd, A.~Arhrib and E.~-M.~Naimi,
  %``Note on tree level unitarity in the general two Higgs doublet model,''
  Phys.\ Lett.\ B {\bf 490}, 119 (2000)
  [hep-ph/0006035].
  %%CITATION = HEP-PH/0006035;%%

\bibitem{Peskin:1991sw}
  M.E.~Peskin and T.~Takeuchi,
  %``Estimation of oblique electroweak corrections,''
  Phys.\ Rev.\ D {\bf 46}, 381 (1992).
  %%CITATION = PHRVA,D46,381;%%

\bibitem{lepewwg}
The ALEPH, CDF,  D0, DELPHI, L3, OPAL, SLD Collaborations, the LEP Electroweak Working Group, the Tevatron Electroweak Working Group, and the SLD electroweak and heavy flavour Groups,
  %``Precision Electroweak Measurements and Constraints on the Standard Model,''
  arXiv:1012.2367 [hep-ex].
  %%CITATION = ARXIV:1012.2367;%%

\bibitem{gfitter1}
  M.~Baak, M.~Goebel, J.~Haller, A.~Hoecker, D.~Ludwig, K.~Moenig, M.~Schott and J.~Stelzer,
  %``Updated Status of the Global Electroweak Fit and Constraints on New Physics,''
  Eur.\ Phys.\ J.\ C {\bf 72}, 2003 (2012)
  [arXiv:1107.0975 [hep-ph]].
  %%CITATION = ARXIV:1107.0975;%%

\bibitem{gfitter2}
M.~Baak, M.~Goebel, J.~Haller, A.~Hoecker, D.~Kennedy, R.~Kogler, K.~Moenig, M.~Schott and J.~Stelzer,
  %``The Electroweak Fit of the Standard Model after the Discovery of a New Boson at the LHC,''
  arXiv:1209.2716 [hep-ph].
  %%CITATION = ARXIV:1209.2716;%%

\bibitem{bphys1}
T.~Hermann, M.~Misiak and M.~Steinhauser,
  %``$\bar{B}\to X_s \gamma$ in the Two Higgs Doublet Model up to
Next-to-Next-to-Leading Order in QCD,''
  JHEP {\bf 1211} (2012) 036
  [arXiv:1208.2788 [hep-ph]].

\bibitem{bphys2}
  F. Mahmoudi, talk given at Prospects For Charged Higgs Discovery At Colliders
(CHARGED 2012), 8-11 October, Uppsala, Sweden.

\bibitem{cms_atlas}
For recent results,
see
%%%ATLAS
%\bibitem{:2012gk}
G.~Aad {\it et al.}  [ATLAS Collaboration],
 %``Observation of a new particle in the search for the Standard Model Higgs
 %boson with the ATLAS detector at the LHC,''
Phys.\ Lett.\ B \textbf{716}, 1 (2012)
[arXiv:1207.7214 [hep-ex]].
%%CITATION = ARXIV:1207.7214;%%
%%
%%CMS
%\bibitem{:2012gu}
S.~Chatrchyan \textit{et al.}  [CMS Collaboration],
 %``Observation of a new boson at a mass of 125 GeV with the CMS experiment
 %at the LHC,''
Phys.\ Lett.\ B \textbf{716}, 30 (2012)
[arXiv:1207.7235 [hep-ex]].
%%CITATION = ARXIV:1207.7235;%%

\bibitem{average}
  A.~Arbey, M.~Battaglia, A.~Djouadi and F.~Mahmoudi,
  %``An update on the constraints on the phenomenological MSSM from the new LHC Higgs results,''
  arXiv:1211.4004 [hep-ph].
  %%CITATION = ARXIV:1211.4004;%%

\bibitem{Moriond}
The ATLAS collaboration, ATLAS-CONF-2013-012, ATLAS-CONF-2013-013,
ATLAS-CONF-2013-030, ATLAS-CONF-2013-034.

The CMS collaboration, CMS notes CMS-PAS-HIG-13-001, CMS-PAS-HIG-13-002,
CMS-PAS-HIG-13-003 and CMS-PAS-HIG-13-003.



\end{thebibliography}

\end{document}